\documentclass[sigconf,natbib=true,anonymous=false]{acmart}
\AtBeginDocument{%
  \providecommand\BibTeX{{%
    \normalfont B\kern-0.5em{\scshape i\kern-0.25em b}\kern-0.8em\TeX}}}

\copyrightyear{2024} 
\acmYear{2024} 
\setcopyright{rightsretained} 
\acmConference[SIGIR '24]{Proceedings of the 47th International ACM SIGIR Conference on Research and Development in Information Retrieval}{July 14--18, 2024}{Washington, DC, USA}
\acmBooktitle{Proceedings of the 47th International ACM SIGIR Conference on Research and Development in Information Retrieval (SIGIR '24), July 14--18, 2024, Washington, DC, USA}
\acmDOI{10.1145/3626772.3657910}
\acmISBN{979-8-4007-0431-4/24/07}

\usepackage{amsmath,flushend,boldline,paralist}
\usepackage{booktabs}
\usepackage{amsfonts}
\usepackage{amsbsy}
\usepackage{bbm}
\usepackage{algorithm}
\usepackage{algorithmic}
\usepackage{bm}
\usepackage{enumitem}
\usepackage{lipsum}
\usepackage{multirow}
\usepackage[normalem]{ulem}

\usepackage{graphicx}
\usepackage{subcaption}

\usepackage{xurl}
\usepackage{tablefootnote}

\usepackage{centernot}
\newtheorem*{problem}{Problem}

\begin{document}

\title{SpherE: Expressive and Interpretable Knowledge Graph Embedding for Set Retrieval}

\author{Zihao Li}
\affiliation{%
  \institution{University of Illinois at Urbana-Champaign}
  \state{Illinois}
  \country{USA}
}
\email{zihaoli5@illinois.edu}
\orcid{0000-0001-7987-1770}

\author{Yuyi Ao}
\affiliation{%
  \institution{University of Illinois at Urbana-Champaign}
  \state{Illinois}
  \country{USA}
}
\email{yuyiao2@illinois.edu}

\author{Jingrui He}
\affiliation{%
  \institution{University of Illinois at Urbana-Champaign}
  \state{Illinois}
  \country{USA}
}
\email{jingrui@illinois.edu}

\renewcommand{\shortauthors}{Zihao Li, et al.}

\begin{abstract}
    Knowledge graphs (KGs), which store an extensive number of relational facts $(head, relation, tail)$, serve various applications. While many downstream tasks highly rely on the expressive modeling and predictive embedding of KGs, most of the current KG representation learning methods, where each entity is embedded as a vector in the Euclidean space and each relation is embedded as a transformation, follow an entity ranking protocol. On one hand, such an embedding design cannot capture \textit{many-to-many} relations. On the other hand, in many retrieval cases, the users wish to get an exact set of answers without any ranking, especially when the results are expected to be precise, e.g., which genes cause an illness. Such scenarios are commonly referred to as ``set retrieval". This work presents a pioneering study on the KG set retrieval problem. We show that the set retrieval highly depends on expressive modeling of many-to-many relations, and propose a new KG embedding model SpherE to address this problem. SpherE is based on rotational embedding methods, but each entity is embedded as a sphere instead of a vector. While inheriting the high interpretability of rotational-based models, our SpherE can more expressively model one-to-many, many-to-one, and many-to-many relations. Through extensive experiments, we show that our SpherE can well address the set retrieval problem while still having a good predictive ability to infer missing facts. The code is available at \url{https://github.com/Violet24K/SpherE}.
\end{abstract}






\maketitle

\section{Introduction}
Knowledge Graphs (KGs), e.g., the widely used YAGO \cite{DBLP:conf/www/SuchanekKW07}, Freebase \cite{DBLP:conf/sigmod/BollackerEPST08}, DBpedia \cite{DBLP:conf/semweb/AuerBKLCI07}, WordNet \cite{DBLP:journals/cacm/Miller95}, have been serving multiple downstream applications such as information retrieval \cite{DBLP:conf/www/XiongPC17}, recommender systems \cite{DBLP:conf/aaai/ZhouLLLG17, DBLP:conf/kdd/ZhangYLXM16}, natural language processing \cite{DBLP:conf/icml/YangCS16, DBLP:conf/acl/YangM17}, multimedia network analysis \cite{DBLP:conf/www/ZhangZYPZHWH17, DBLP:journals/pami/YanXZZYL07}, question answering ~\cite{binet,prefnet}, fact checking ~\cite{kompare,inspector}. To utilize the extensive amount of knowledge in the KG, many works have studied Knowledge Graph Embedding (KGE), which learns low-dimensional representations of entities and relations of them \cite{DBLP:journals/tnn/JiPCMY22, DBLP:journals/tkdd/RossiBFMM21, DBLP:journals/symmetry/WangQW21, DBLP:conf/cikm/XiongNDC23, DBLP:journals/tkde/WangMWG17}. Starting from TransE \cite{DBLP:conf/nips/BordesUGWY13}, a group of translation-based methods TransH \cite{DBLP:conf/aaai/WangZFC14}, TransR \cite{DBLP:conf/aaai/LinLSLZ15}, TransD \cite{DBLP:conf/acl/JiHXL015}, TorusE \cite{DBLP:conf/aaai/EbisuI18} model the relation as translations between entities in the embedding space. However, the translation-based methods suffer from the inexpressiveness when modeling \textit{symmetric relations}\footnote{A relation $r$ is symmetric if $\forall x,y$, $r(x,y) \Rightarrow r(y,x)$.} because translation itself is an anti-symmetry transformation. In recent years, some seminal works have proposed to model the relations by rotation, such as RotatE \cite{DBLP:conf/iclr/SunDNT19}, QuatE \cite{DBLP:conf/nips/0007TYL19}, Rotate3D \cite{DBLP:conf/cikm/GaoSS0W20}. The rotation-based methods can inherently handle symmetry and anti-symmetry relations\footnote{$k\pi$-rotation is symmetric since a circle is $2k\pi$.}. However, all the methods that model each entity as a vector (i.e., a point) and each relation as a single function mapping or injective mapping, are inexpressive when modeling \textit{one-to-many, many-to-one}, and \textit{many-to-many} relations\footnote{This is because, if $f$ is a function mapping or injective mapping, then either $f(x)$ or $f^{-1}(x)$ is unique
% \he{not both?}
, hence the \textit{many} side cannot be expressively modeled.}. Later on, BoxE \cite{DBLP:conf/nips/AbboudCLS20} and HousE \cite{DBLP:conf/icml/Li0LH0LSWDSXZ22} are proposed by more complex modeling of the entities and relations. Though these two methods are proven to have high expressiveness, the complex modeling makes the model less interpretable, and the real-world meanings of the parameters haven't been fully discussed. 
% For BoxE, embedding the relations as hyper-rectangles instead of a transformation is counter-intuitive. For HousE, an interpretation of the Householder projections has not yet been discussed.
% \he{Consider removing the previous two sentences -- a little too harsh.} 
Bilinear models, such as RESCAL \cite{DBLP:conf/icml/NickelTK11}, DistMult \cite{DBLP:journals/corr/YangYHGD14a}, ComplEx \cite{DBLP:conf/icml/TrouillonWRGB16} and SimplE \cite{DBLP:conf/nips/Kazemi018}, aim to capture relations as bilinear products between the vector embeddings of entities and relations. Some methods also try to use neural network architectures for KGE, such as SME \cite{DBLP:journals/corr/abs-1301-3485}, NTN \cite{DBLP:conf/nips/SocherCMN13}, MLP \cite{DBLP:conf/kdd/0001GHHLMSSZ14} and NAM \cite{DBLP:journals/corr/Liu0LWH16}.

Furthermore, current KGE research mostly follows an entity ranking protocol. When we ask the model to predict the tail entity of the triple $(head, relation, ?)$, or the head entity of the triple $(?, relation, tail)$, the model assigns every possible triple a score to measure the plausibility
of its existence, and returns a ranked list of all the entities \cite{DBLP:conf/sigir/ZhouCHY022} based on the score of each completed triple. However, given the head entity and the relation, there might be more than one correct tail entities. In some real scenarios where the user requires an exact set of correct answers, determining the appropriate threshold to cut off the ranked list poses a challenge. For instance, in querying a bioinformatic knowledge graph to identify genes associated with an illness, the abundance of potential genes complicates the threshold determination process significantly. To this end, we introduce and formulate a new problem, named \textit{Knowledge Graph Set Retrieval}, which aims to exactly find all the entities of the triple queries $(head, relation, ?)$ or $(?, relation, tail)$, given an incomplete knowledge graph. We propose SpherE that embeds each entity as a sphere instead of a vector in the Euclidean space, and each relation as a rotation. Based on the dimension of rotation, SpherE consists of a group of methods SpherE-$k$D ($k\geq 2$). The intuition of embedding entities as spheres lies in the modeling of ``\textit{universality}" of each entity, which enhances the interpretability of SpherE. We validate such intuition through experiments. We also show that SpherE is expressive in modeling many relations patterns (Table \ref{tb: inference patterns}) and demonstrate that our SpherE outperforms the baseline methods on the knowledge graph set retrieval task.

\begin{table}[t]
\vspace{-4mm}
\caption{SpherE can expressively model the inference patterns that state-of-the-art methods are able to model.}
% \he{comparison with SOTA?}
\label{tb: inference patterns}
\vspace{-3mm}
% \vskip 0.15in
\begin{center}
\begin{small}
% \begin{sc}
\scalebox{0.70}{
\begin{tabular}{lcccr}
\toprule
Inference pattern\tablefootnote{Symmetry: $r(x,y) \Rightarrow r(y,x)$; Anti-symmetry: $r(x,y) \Rightarrow \neg r(y,x)$; Inversion: $r_1(x,y) \Leftrightarrow r_2(y,x)$; Composition: $r_1(x,y) \wedge r_2(y,z) \Leftrightarrow r_3(x,z)$.} & SpherE-2D & SpherE-3D & SpherE-$k$D $(k \geq 4)$ \\
\midrule
Symmetry        & $\surd$& $\surd$& $\surd$ \\
Anti-symmetry   & $\surd$& $\surd$& $\surd$ \\
Inversion      & $\surd$& $\surd$& $\surd$ \\
Composition  & $\surd$& $\surd$& $\surd$ \\
NC composition\tablefootnote{Non-commutative composition \cite{DBLP:conf/cikm/GaoSS0W20}. Let $r_3 = r_1 \circ r_2$. $r_1(x,y) \wedge r_2(y,z) \Leftrightarrow r_3(x,z)$, but $(r_1 \circ r_2)(x, y) \centernot\Rightarrow(r_2 \circ r_1)(x, y)$.}     & $\times$& $\surd$& $\surd$ \\
Multiplicity\tablefootnote{there exists multiple relations satisfying $r(x, y)=true$.}    & $\times$& $\surd$& $\surd$ \\
RMPs\tablefootnote{RMPs means relation mapping properties, containing one-to-one, one-to-many, many-to-one and many-to-many. Some works may use "n" for "many".}     & $\surd$& $\surd$& $\surd$ \\
% many-to-one     & $\surd$& $\surd$& $\surd$ \\
% many-to-many    & $\surd$& $\surd$& $\surd$ \\
\bottomrule
\end{tabular}
}
% \end{sc}
\end{small}
\end{center}
\vskip -0.1in
\vspace{-3mm}
\end{table}

% \vspace{-3mm}
\section{Preliminary}
\textbf{Knowledge Graph.} A knowledge graph $\mathcal{G} = (\mathcal{E}, \mathcal{R}, \mathcal{T})$ is defined by an entity set $\mathcal{E}$, a relation set $\mathcal{R}$ and a triple set $\mathcal{T}$. A triple $(h, r, t) \in \mathcal{T}$ is defined by the head entity $h$, tail entity $t$ and their relation $r$. We use $r(h, t) \in \{true, false\}$ to denote whether the fact $(h, r, t)$ holds in the real world,  i.e., whether $(h, r, t)$ is a fact in a complete and correct oracle knowledge graph or not. 

\noindent\textbf{Tail/Head Query.} A tail query $(h, r, ?)$ aims to find all $t\in \mathcal{G}$ such that $r(h, t)$ is $true$, and similarly a head query $(?, r, t)$ aims to find all $h\in \mathcal{G}$ such that $r(h, t)$ is $true$. In the real world and real KGs, it is possible that a head or tail query has more than one answers. For example, the query $(Max\_Born, advises, ?)$ can have correct tails in the KG to be $Robert\_Oppenheimer, Pascual\_Jordan$, or $Carl\_Hermann$.

\noindent\textbf{Knowledge Graph Embedding (KGE).} For an arbitrary triple $(h, r, t)$, assume the embedding of $h$ and $t$ are respectively $\bm{h}$, $\bm{t}$, and the transformation mapping of relation $r$ is $f_r(\cdot)$. We expect that $r(h, t) \iff \bm{t} = f_r(\bm{h})$.

When testing the performance KGE methods, a widely accepted and used benchmark protocol is, for each triple in the test dataset, we mask the tail and pass a tail query $(h, r, ?)$ to the model. Then, the model computes the score of $(h, r, e)$ for all the entities $e\in \mathcal{E}$ and returns a ranked list. Then, we check where the masked tail is ranked in the list, and do a similar test by masking heads.
In this paper, we introduce and formulate another related but different problem: the knowledge graph set retrieval problem.
\begin{problem}{Knowledge Graph Set Retrieval.}
% \he{Please provide the intuitive meaning of $F1$.}
% \he{This should be named Problem Definition?}
\begin{description}
\item[Input:] 
(i) a knowledge graph $\mathcal{G} = (\mathcal{E}, \mathcal{R}, \mathcal{T})$\\
(ii) A relation $r$. For tail-set retrieval, a head $h$; for head-set retrieval, a tail $t$. Equivalently, a head/tail query.
\item[Output:] For tail-set retrieval, all the $h \in \mathcal{E}$ such that $r(h, t)$ is $true$. For head-set retrieval, all the $t \in \mathcal{E}$ such that $r(h, t)$ is $true$
\end{description}
\end{problem}

\begin{figure}[t]
    \centering
    \includegraphics[width=0.47\textwidth]{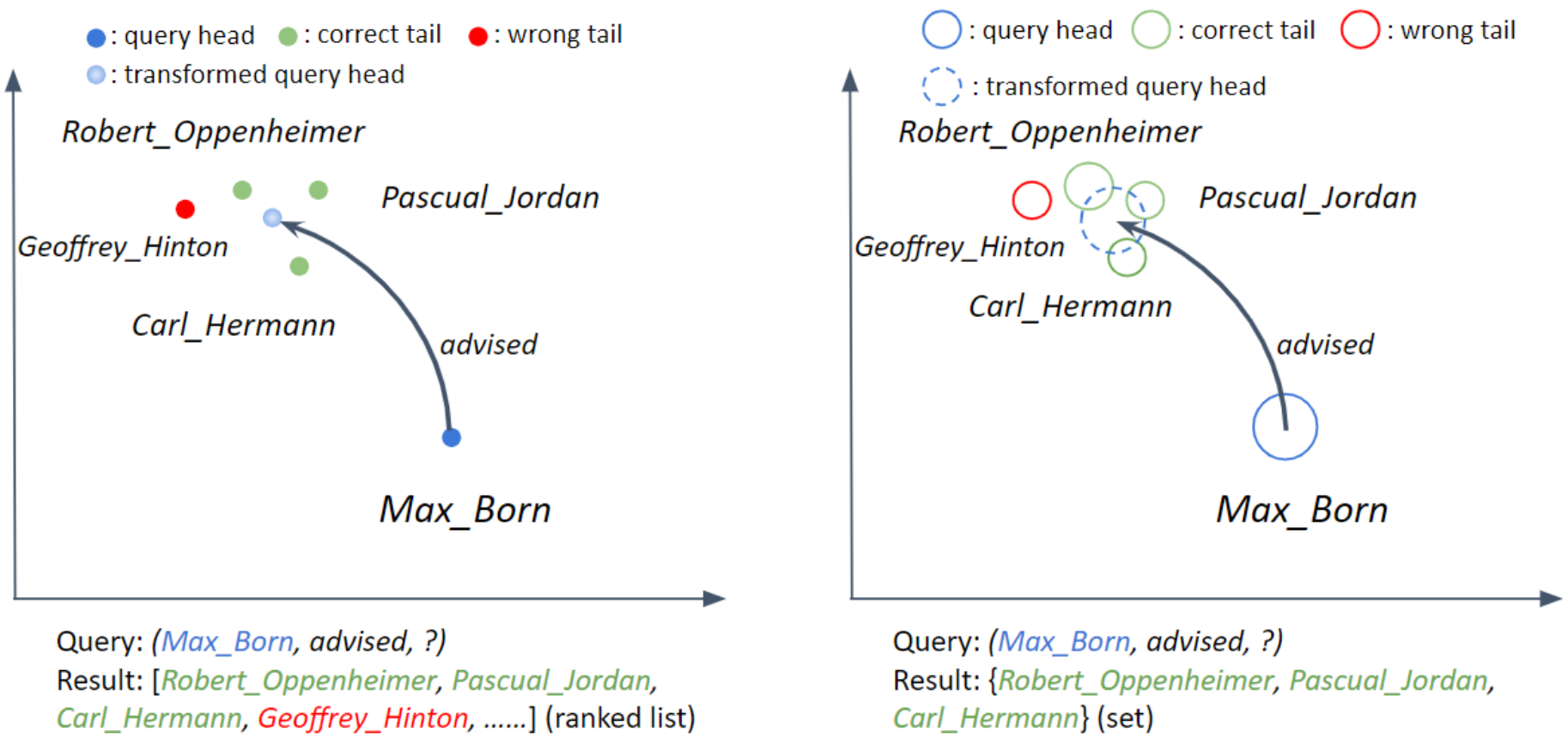}
    \vspace{-4mm}
    \caption{Illustration of Rotational-based embedding methods (left) and SpherE (right) in 2D Euclidean space. The user asks the tail query $(Max\_Born, advised, ?)$. In the KG, $Robert\_Oppenheimer, Pascual\_Jordan$, or $Carl\_Hermann$ are correct answers to this query, but $Geoffrey\_Hinton$ is not. While the left returns a ranked list of all entities in the KG and disproportionately ranks incorrect answers highly, our SpherE method returns a set that exclusively comprises all correct tail entities. The radius of each sphere is learned.} 
    \label{fig: illustration}
\vspace{-4mm}
\end{figure}

To address the knowledge graph set retrieval problem, we propose our SpherE with an illustration in Figure \ref{fig: illustration}. SpherE embeds the entities as spheres instead of vectors and models the triple $(h, r, t)$ using \textit{non-disjoint\footnote{Tangent, intersecting, concentric or contained all considered non-disjoint.} $h$ and $t$ spheres}. In particular, $r(h, t)$ is $true$ if the ball bounded by the transformed $h$ sphere overlaps with the ball bounded by the $t$ sphere. SpherE can naturally express the one-to-many, many-to-one, and many-to-many relations, and enable knowledge graph set retrieval. The interpretation of entity radius lies in the universality of an entity, i.e., how often the entity appears in the KG triples. For example, ``apple" (both fruit and brand) is very common in daily life, and there might be many facts with it as head or tail. In this case, the radius of ``apple" should be large to make the apple sphere be non-disjoint with other related spheres.

\section{Proposed Method}
\subsection{Sphere-based Modeling}
SpherE is based on the rotational embedding methods RotatE \cite{DBLP:conf/iclr/SunDNT19}, Rotate3D \cite{DBLP:conf/cikm/GaoSS0W20}, and HousE\footnote{More precisely, HousE-r, where rotations are modeled by Householder matrices .} \cite{DBLP:conf/icml/Li0LH0LSWDSXZ22}. They respectively embed the relations into 2D, 3D, and $k$D Euclidean spaces ($k \in \mathbb{N}_{+}$). In SpherE, each relation is also embedded as a rotation, but each entity is embedded as a sphere. The rotations in any dimension, represented by rotation matrix multiplication, are invertible because of the provably invertible  rotation matrices. Therefore, both the tail query and head query can be handled equivalently. For an entity $e$, we denote the center and the radius of $e$'s sphere embedding to be $\bm{c}_e$ and $r_e$. When a tail query $(h, r, ?)$ is passed to SpherE, for all $t\in \mathcal{E}$, it checks whether the relational transformed $h$ sphere and $t$ sphere intersect. If so, $t$ is included in the set to return. In other words, we expect
\begin{equation}
\label{eq: htiff1}
    r(h, t) \iff \|f_r(\bm{c}_h) - \bm{c_t}\|_2 \leq r_h + r_t
\end{equation}
Similarly, when facing a head query, we expect
\begin{equation}
\label{eq: htiff2}
    r(h, t) \iff \|\bm{c}_h - f^{-1}_r(\bm{c_t})\|_2 \leq r_h + r_t
\end{equation}

At first glance, Equations \ref{eq: htiff1} and \ref{eq: htiff2} might look contradictory because by combining them together, the model must satisfy:
\begin{equation}
\label{eq: htiff3}
    \|f_r(\bm{c}_h) - \bm{c_t}\|_2 \leq r_h + r_t \iff \|\bm{c}_h - f^{-1}_r(\bm{c_t})\|_2 \leq r_h + r_t
\end{equation}
In fact, Equation \ref{eq: htiff3} holds for SpherE because the rotation transformation and its inverse are isometric, i.e., distance-preserving, and we do have
\begin{equation}
\|\bm{c}_h - f^{-1}_r(\bm{c_t})\|_2 = \|f_r(\bm{c}_h) - f_r(f^{-1}_r(\bm{c_t}))\|_2 = \|f_r(\bm{c}_h) - \bm{c_t}\|_2 
\end{equation}

We theoretically show that our modeling is expressive to model all the inference patterns in Table \ref{tb: inference patterns}. Due to the space limit, we only show the proof on the high level.

\begin{theorem}
\label{thm: sicr}
    SpherE can model symmetry, inversion, and composition relation patterns.
\end{theorem}
\begin{proof}
    It has been proved that RotatE, RotatE3D, and HousE can model these relation patterns. By taking the embedding parameters of these models, for each $r(h, t)$ that is successfully modeled by RotatE, RotatE3D or HousE, $f_r(c_h) = c_t$. Pick $r_e > 0$ for all entities, we have $\|f_r(\bm{c}_h) - \bm{c_t}\|_2 \leq r_h + r_t \iff r(h, t)$. 
\end{proof}

\begin{theorem}
    SpherE-$k$D ($k\geq 3$) can model anti-symmetry, non-commutative composition, and multiplicity inference patterns. SpherE-2D can model anti-symmetry
\end{theorem}
\begin{proof}
    Similar to Theorem \ref{thm: sicr}, but pick $r_e \to 0_+$.
\end{proof}

Similar to RotatE, SpherE-2D cannot model the non-commutative composition and multiplicity inference patterns. 

\begin{theorem}
    SpherE can model the relation mapping properties.
\end{theorem}
\begin{proof}
    As shown in Figure \ref{fig: illustration}, given a head entity and a relation, multiple tail spheres can be non-disjoint with the rotated head entity. Therefore SpherE can model one-to-many relations. Similarly, SpherE can also model many-to-one relations. Combining these together, SpherE can model the relation mapping properties.
\end{proof}

\vspace{-3mm}
\subsection{Optimization}
We illustrate the optimization method of the model. The parameters to be learned are the rotational relation embeddings $f_r$ for each relation, the sphere center $c_e$ of each entity $e$, and the sphere radius $r_e$ of each entity $e$. Respectively, for SpherE-2D, SpherE-3D, and Sphere-$k$D, we use the same negative sampling strategy as in RotatE, RotatE3D, and HousE. 
% Assume $\mathcal{T}$ is the set of all positive triples in the training set, and $\mathcal{T}_n$ to be the set of negative triples. Following the sigmoid loss for rotational embedding, we design our loss function as 
For a triple $x = (h, r, t)$, let $\mathcal{T}_n$ be the set of its negative triples. Following the sigmoid loss for rotational embedding, we design our loss function as

\begin{equation}
\begin{split}
    \mathcal{L} = -\log&\sigma(\gamma - d(x)) - \sum_{x' \in \mathcal{T}_n}p(x')\log\sigma(d(x') -\gamma)\\
    % \mathcal{L} &= \sum_{x \in \mathcal{T}} \sum_{x' \in \mathcal{T}_{n}} [\gamma + d(x) - d(x')]_+   \\
    % x &= (h, r, t), x' = (h', r, t')\\
    d(h, r, t) &= [||f_r(\bm{c_h}) - \bm{c_t}||_{p} -  r_h - r_t - \alpha r_h - \beta r_t]_+\\
\end{split}
\end{equation}
where $\alpha\geq 0$, $\beta \geq 0$ are hyperparameters close to $0$ to encourage sphere intersections, and $[x]_+ = ReLU(x)$. $p=2$ defines the norm to use. By such an objective function, during optimization, for positive triples $(h, r, t)$, the trainer encourages $||f_r(\bm{c_h}) - \bm{c_t}||_{p} \leq r_h + r_t$, but does not encourage $||f_r(\bm{c_h}) - \bm{c_t}||_{p}$ to be to small after the transformed head sphere and tail sphere become non-disjoint; for a negative sample $(h', r, t')$, the learner encourages $||f_r(\bm{c_h}) - \bm{c_t}||_{p}$ to be as large as possible, i.e., encouraging the transformed head to be disjoint with the tail sphere. We demonstrate the effectiveness of our optimization method and the expressive of SpherE modeling by extensive experiments on knowledge graph set retrieval.

\section{Experiments}
\subsection{Experimental Setup}
\subsubsection{Datasets}
We use two widely-used knowledge graph embedding benchmark datasets, FB15K237 and WN18RR, in our experiments. Table \ref{TB: statistics of KGs} shows their statistics.
\begin{table}[h]
\vspace{-2mm}
\centering
\caption{Statistics of Datasets}
\vspace{-4mm}
\label{TB: statistics of KGs}
\begin{small}
\begin{tabular}{|l|lll|}
\hline
KG      & $|\mathcal{E}|$  & $|\mathcal{R}|$     & \# train/valid/test \\ \hline\hline
FB15K237    & 14,541 & 237 & 272,115/17,535/20,446 \\ \hline
WN18RR & 40,943 & 11   & 86,835/17,535/20,446 \\ \hline
\end{tabular}
\end{small}
\vspace{-4mm}
\end{table}

\subsubsection{Baselines}
To the best of our knowledge, this paper is the first to study knowledge graph set retrieval, so we convert the state-of-the-art knowledge graph embedding methods RotatE, RotatE3D, and HousE, to handle knowledge graph set retrieval settings. For each of the methods, given a tail/head query, it returns a ranked list of the entities. We truncate the ranked list to get the top-$l$ set, which contains the $l$ elements ranked in the front of the list. We tried many choices of $l$ and reported the performance of $l=1, 3, 5, 10, 20, 100$ for comparison. Experimentally, respectively for FB15K237 and WN18RR, the baseline models have the best performance when $l=10$ and $l=3$.

\subsubsection{Settings and Metrics}
We first train each method to near convergence. Then, for each triple $(h, r, t)$ in the test dataset, we respectively mask the head and tail of it to produce two queries $(h, r, ?)$ and $(?, r, t)$. Take $(h, r, ?)$ as an example: we then feed the query into the model and get a set of retrieved tails $\mathcal{E}_m$. Then, we compute the ground-truth tails $\mathcal{E}_g$ from the training data, valid data, and test data. We calculate the F1 score of $\mathcal{E}_m$ and $\mathcal{E}_g$ as a measurement of knowledge graph set retrieval performance. We also measure the inferring ability of the model by computing the probability that $t$, the actual tail of $(h, r, t)$, is in the set retrieved from $(h, r, ?)$. We record this probability as ``Retrieve Rate". We repeat a similar procedure for the query $(?, r, t)$.

% \subsubsection{Reproducibility} 
% We use $p=2$, $\alpha=0.1, \beta=0$ for head queries, $p=2$, $\alpha=0, \beta=0.1$ for tail queries.

\begin{table}[t]
\vspace{-3mm}
\setlength{\tabcolsep}{0.9mm}{
\caption{Comparison($\uparrow$) of Embedding Methods.}
\label{tb: fb15k237 results}
\begin{center}
% \vskip 2mm
\vspace{-3mm}
\begin{small}
\scalebox{1}{
\begin{tabular}{l|l|cc|cc|c}
\toprule
\multicolumn{2}{c|}{\multirow{2}{*}{Models}} & \multicolumn{5}{c}{FB15K237}\\
\multicolumn{2}{c|}{}               & Head F1 &  Tail F1 & Head RR & Tail RR & n-to-n F1\\
\midrule
        & top-1     & 0.132& 0.281& 0.044& 0.174 &0.158\\
        & top-3     & 0.212& 0.346& 0.092& 0.280 &0.283\\
        & top-5     & 0.245& 0.361& 0.133& 0.344 &0.333\\
RotatE  & top-10    & 0.270& 0.359& 0.202& 0.442 &0.367\\
        & top-20    & 0.276& 0.332& 0.297& 0.554 &0.364\\
        & top-100   & 0.246& 0.216& 0.513& 0.766 &0.275\\
        & Sphere    & \textbf{0.451}& \textbf{0.444}& 0.761& 0.760 &\textbf{0.447}\\
\midrule
        & top-1     & 0.132& 0.287& 0.046& 0.188 &0.157\\
        & top-3     & 0.210& 0.345& 0.097& 0.289 &0.281\\
        & top-5     & 0.242& 0.359& 0.140& 0.354 &0.331\\
RotatE3D& top-10    & 0.269& 0.356& 0.214& 0.454 &0.365\\
        & top-20    & 0.274& 0.329& 0.308& 0.564 &0.363\\
        & top-100   & 0.246& 0.216& 0.527& 0.778 &0.275\\
        & Sphere    & \textbf{0.444}& \textbf{0.430}& 0.779& 0.779 &\textbf{0.462}\\
\midrule
        & top-1     & 0.051& 0.192& 0.043& 0.177 &0.121\\
        & top-3     & 0.082& 0.207& 0.095& 0.281 &0.145\\
        & top-5     & 0.102& 0.215& 0.137& 0.346 &0.159\\
HousE-5D & top-10    & 0.129& 0.217& 0.211& 0.444 &0.174\\
        & top-20    & 0.145& 0.210& 0.294& 0.540 &0.178\\
        & top-100   & 0.137& 0.152& 0.476& 0.743 &0.145\\
        & Sphere    & \textbf{0.384}& \textbf{0.345}& 0.622& 0.653 &\textbf{0.366}\\
\midrule
        & top-1     & 0.060& 0.208& 0.051& 0.191 &0.134\\
        & top-3     & 0.098& 0.229& 0.114& 0.304 &0.162\\
        & top-5     & 0.120& 0.134& 0.157& 0.375 &0.179\\
HousE-10D& top-10    & 0.150& 0.246& 0.239& 0.485 &0.197\\
        & top-20    & 0.169& 0.239& 0.332& 0.588 &0.205\\
        & top-100   & 0.173& 0.174& 0.539& 0.793 &0.175\\
        & Sphere    & \textbf{0.401}& \textbf{0.412}& 0.539& 0.573 &\textbf{0.409}\\
\midrule
        & top-1     & 0.059& 0.190& 0.051& 0.191 &0.134\\
        & top-3     & 0.096& 0.228& 0.111& 0.304 &0.162\\
        & top-5     & 0.119& 0.238& 0.156& 0.373 &0.179\\
HousE-20D& top-10    & 0.151& 0.244& 0.238& 0.482 &0.198\\
        & top-20    & 0.171& 0.238& 0.322& 0.588 &0.205\\
        & top-100   & 0.174& 0.174& 0.540& 0.793 &0.175\\
        & Sphere    & \textbf{0.378}& \textbf{0.406}& 0.463& 0.495& \textbf{0.394}\\
\bottomrule
\end{tabular}
}
\end{small}
\end{center}
}
\vspace{-4mm}
\end{table}

\vspace{-1mm}
\subsection{Experiment Results}
We use $p=2$, $\alpha=0.1, \beta=0$ for head queries, $p=2$, $\alpha=0, \beta=0.1$ for tail queries. The main experiment results are shown in Table \ref{tb: fb15k237 results} and \ref{tb: wn18rr results}. Due to the space limitation, we show full experiment data only for the FB15K237 dataset. For each RotatE, RotatE3D and HousE-$k$D, each top-$l$ row stands for a top-$l$ set, and the Sphere row stands for our SpherE-2D, SpherE-3D and SpherE-$k$D, respectively. The Head/Tail F1 stands for the average F1 score of head/tail queries, and the Head/Tail RR stands for the average retrieve rate of the correct head/tail of the test triple. The n-to-n F1 stands for the average F1 score of querying \textit{many-to-many} relations. We mark the best performance in each base method (RotatE, RotatE3D, or HousE-$k$D) category in bold.

First, our SpherE outperforms all the baseline methods in terms of both Head F1 and Tail F1 on both datasets. Second, after embedding the entities as spheres instead of vectors, even though we still embed the relations to be rotations, the performance on the knowledge graph set retrieve task is boosted significantly, which validates the expressiveness of our SpherE model. Third, SpherE significantly outperforms the baseline methods in terms of n-to-n F1, which shows the superior ability of SpherE in modeling \textit{many-to-many} relations. Fourth, SpherE still has a good ability to infer the missing links in the KG, as the Head RR and Tail RR of unseen triples are comparable with the top-20 set of FB15K237 and the top-3 set of WN18RR. On average, each head/tail query of FB15K237 has 3 or 5 correct heads/tails, and each head/tail query of WN18RR has 1 or 3 correct heads/tails. Therefore, the top-20 set of FB15K237 and the top-3 set of WN18RR are strong baselines for inferring the correct entity. Fifth, we observe that, as the dimension of rotation increases, the algorithms perform worse on knowledge graph set retrieval tasks. It could be a future direction to study what causes the performance drop.

We also validate the interpretability of the radius through experiments. As discussed previously, the radius of an entity encodes its universality, i.e., how often it appears in the KG triples. We validate this by calculating the average radius of the entities that have the same number of appearances in the train, valid and test triples. The results show that, for both datasets, the more time an entity occurs, the larger the radius is optimized. Interestingly, when an entity appears only 1 time in the KG, it is even optimized to have a negative radius to make its sphere disjoint with others.

\begin{table}[t]
\vspace{-3mm}
\setlength{\tabcolsep}{0.9mm}{
\caption{Comparison($\uparrow$) of Embedding Methods.}
\label{tb: wn18rr results}
\begin{center}
% \vskip 2mm
\begin{small}
\vspace{-3mm}
\scalebox{1}{
\begin{tabular}{l|l|cc|cc|c}
\toprule
\multicolumn{2}{c|}{\multirow{2}{*}{Models}} &\multicolumn{5}{c}{WN18RR} \\
\multicolumn{2}{c|}{}               & Head F1 &  Tail F1& Head RR & Tail RR & n-to-n F1 \\
\midrule
        & top-1     &0.363& 0.371& 0.147& 0.210 &0.618\\
        & top-3     & 0.458& 0.397& 0.329& 0.431 &0.710\\
        & top-5     &0.435& 0.342& 0.397& 0.512 &0.605\\
RotatE  & top-10    &0.362& 0.240& 0.455& 0.569 &0.400\\
        & top-20    &0.279& 0.155& 0.497& 0.608 &0.232\\
        & top-100   &0.142& 0.052& 0.608& 0.706 &0.053\\
        & Sphere    &\textbf{0.712}& \textbf{0.447}& 0.385& 0.385 &\textbf{0.873}\\
\midrule
        & top-1     & 0.356& 0.360& 0.149& 0.210 &0.612\\
        & top-3     & 0.460& 0.397& 0.336& 0.436 &0.708\\
        & top-5     & 0.439& 0.344& 0.412& 0.521 &0.605\\
RotatE3D& top-10    & 0.365& 0.244& 0.475& 0.592 &0.401\\
        & top-20    & 0.281& 0.158& 0.524& 0.643 &0.231\\
        & top-100   & 0.143& 0.053& 0.643& 0.753 &0.054\\
        & Sphere    &\textbf{0.656}& \textbf{0.426}& 0.413& 0.412 &\textbf{0.779}\\
\bottomrule
\end{tabular}
}
\end{small}
\end{center}
}
\vspace{-3mm}
\end{table}

\begin{figure}[t]
\centering
\begin{subfigure}[b]{0.35\textwidth}
   \includegraphics[width=1\linewidth]{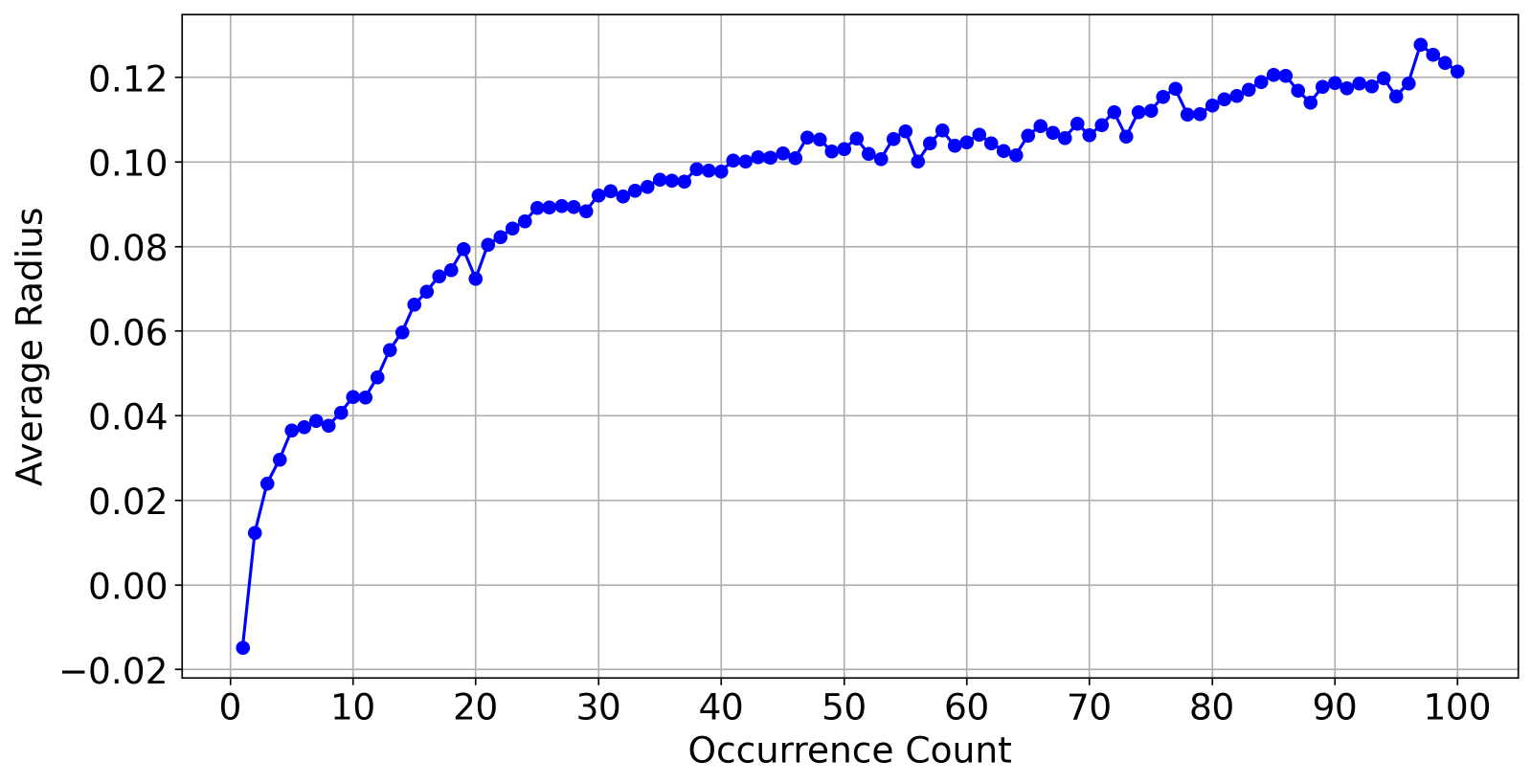}
   \caption{FB15K237}
   \label{fig:Ng1} 
\vspace{-1mm}
\end{subfigure}

\begin{subfigure}[b]{0.35\textwidth}
   \includegraphics[width=1\linewidth]{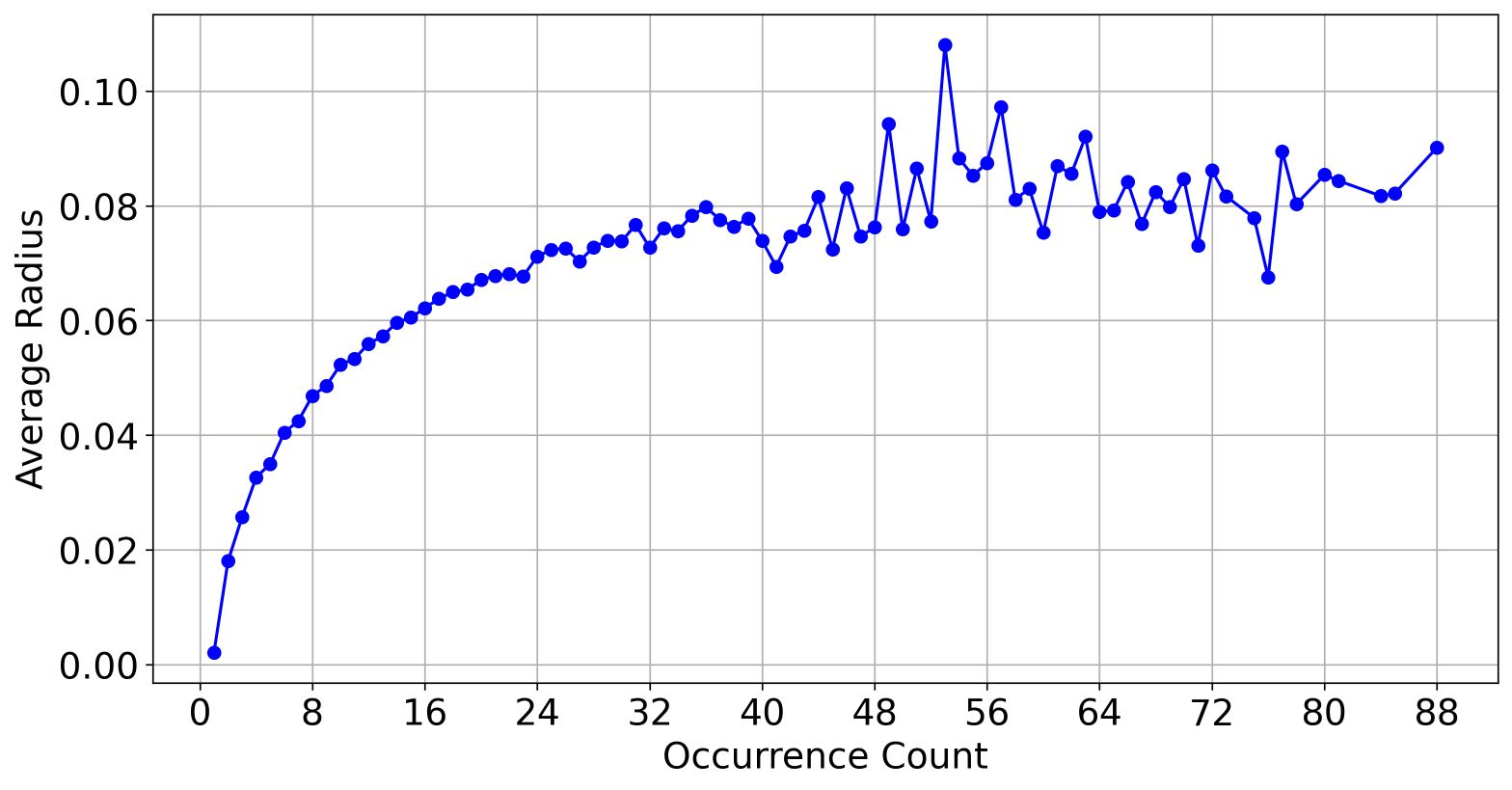}
   \caption{WN18RR}
   \label{fig:Ng2}
\vspace{-1mm}
\end{subfigure}
\vspace{-2mm}
\caption[]{The relation between radius and occurrence count.}
\vspace{-4mm}
\end{figure}

\section{Conclusion}
In this paper, to the best of our knowledge, we first introduce and formulate the problem of knowledge graph set retrieval. We propose SpherE, an expressive and interpretable model to address this problem and demonstrate its effectiveness through extensive experiments on knowledge graph set retrieval.

\clearpage

\clearpage

\section{Acknowledgement}
This work is supported by National Science Foundation under Award No. IIS-2117902. The views and conclusions are those of the authors and should not be interpreted as representing the official policies of the funding agencies or the government.

\bibliographystyle{ACM-Reference-Format}
\bibliography{reference.bib}

\end{document}